\documentclass[12pt]{article}

\usepackage{graphicx}
\usepackage[utf8x]{inputenc}

\usepackage{amssymb}
\usepackage{amsmath}
\usepackage{amsthm}
\newtheorem{definition}{Definition}[section]
\newtheorem{lemma}{Lemma}[section]
\newtheorem{theorem}{Theorem}[section]

\usepackage[a4paper]{geometry}

\newcommand{\R}{\mathbf{R}}

\newcommand{\real}{\operatorname{Re}}
\newcommand{\im}{\operatorname{Im}}

\begin{document}


\title{Creation of higher-energy superposition quantum states motivated by geometric transformations
}
\author{Dimitris Vartziotis\footnotemark[2]\ \footnotemark[3]\ \footnotemark[5] \and Benjamin Himpel\footnotemark[2] \and Markus Pfeil\footnotemark[2]\ \footnotemark[4]}

\maketitle

\renewcommand{\thefootnote}{\fnsymbol{footnote}}
\footnotetext[2]{TWT GmbH Science \& Innovation, Department for Mathematical Research,
  Ernsthaldenstraße 17,
  70565~Stuttgart, Germany}
\footnotetext[3]{NIKI Ltd.\ Digital Engineering, Research Center,
  205~Ethnikis Antistasis Street, 45500~Katsika, Ioannina, Greece}
\footnotetext[4]{Since 9/2017 at Hochschule Ravensburg-Weingarten, Postfach 30 22, 88216 Weingarten, Germany}
\footnotetext[5]{Corresponding author. E-mail address: dimitris.vartziotis@nikitec.gr}
\renewcommand{\thefootnote}{\arabic{footnote}}

\begin{abstract}
 We suggest a way to produce higher-energy superposition states in a circular system of quantum wells. This is inspired by a link to convergence results for geometric transformations of polygons using circulant Hermitian matrices.
\end{abstract}

\section{Introduction}

History shows that developments in mathematics have had an impact on developments in physics and vice versa. Furthermore, non-rigorous results in quantum mechanics have provided geometers and topologists with a lot of questions to answer, while physicists use differential geometry to describe the universe. We present a link between convergence results for geometric transformations of polygons and the dynamical behavior of quantum systems.

Recently, the preparation of specific superposition or entangled state has become more and more important in the domains of quantum information processing \cite{Williams2010,Bouwmeester2000physicsquantuminformation} and also in measuring with quantum systems \cite{Huelga1997ImprovementFrequencyStandards, Leibfried2004TowardHeisenbergLimited}. This is often realized using specific symmetry in Hamilton operators in preparation of the system. 

In particular, we are interested in operators of circulant symmetry. In physical systems these appear typically in systems of circular symmetry (such as closed loops) or, more generally in systems with periodic boundary conditions. While periodic boundary conditions have long been used as a tool for the solid state physicist to approximate extended systems by smaller systems with periodic symmetry, recently there has been an increase interest in truly circulant quantum systems. These systems have been realized as rings of quantum wells in a 2-D plane, which couple only to next neighbors \cite{Unanayan2007}, abstract quantum graphs \cite{Kendon2011PerfectStateTransfer} or as circulant systems of spin qubits \cite{Basic2014Whichweightedcirculant} to name a selection. Other authors begin to investigate efficient quantum circuits to calculate states in such systems \cite{Zhou2016Efficientquantumcircuits}.

We approach the subject from a geometrical perspective. Vartziotis et al \cite{VartziotisWipper2010} have derived a polygon transformation which is characterized by a circulant Hermitian operator. The operator acts on a closed graph (a polygon) in such a way that for repeated application of the operation the polygon is transformed into an eigenshape. Which of the possible eigenshapes the transformation converges on is determined by a single parameter. Furthermore the transformation converges robustly with respect to this parameter. 

Transferring these properties onto a quantum system means that, given the symmetry of the Hamiltonian of the system, a predetermined eigenstate can be reached through time evolution of the system. Unanyan et al \cite{Unanayan2007} have used this approach to achieve time evolution into a desired superposition or entanglement state through adiabatic changes of the operator. Our aim is to show the correspondence of this approach to the geometric algorithm by Vartziotis et al. Readers interested in the dynamics of linear systems of equations are also invited to look at \cite{VartziotisBohnet2014b}, where Vartziotis et al establish the existence of a local attractor which coincides with the set of regular tetrahedra, and which is in spirit similar to the content of our work.

\section{Quantum wells and polygons}

Unanyan et al \cite{Unanayan2007} consider a ring of $n$ quantum wells, the matrix corresponding to the Hamiltonian acting on the $n$ lowest energy eigenstates of the $n$ individual quantum wells and make use of the important property of circulant matrices, that the eigenstates do not depend on the matrix elements of the Hamiltonian, although the eigenvalues do. What this means is that the eigenstates remain in an eigenstate, as long as the Hamilton changes smoothly. This matrix is Hermitian. In fact, Unanyan et al \cite{Unanayan2007} assume without loss of generality that this matrix is symmetric. By changing the distances of the quantum wells as well as their depth, the authors slowly introduce tunneling without changing the symmetry of the system. Therefore, when the systems starts out with an atom in a single well, it will get distributed over time among all wells.

Vartziotis et al \cite{VartziotisWipper2010} have also analyzed circulant Hermitian matrices in order to study a geometric polygon transformation. It turned out that each polygon can be written as a sum of eigenpolygons and that repeated application of the transformation will transform the polygon into a chosen eigenpolygon. The fact that the polygon is transformed into an eigenpolygon corresponds to the aim of transforming an arbitrary state of the system into an eigenstate. The fact that the transformation is a circulant Hermitian matrix suggests that the transformation process can be viewed as a kind of Hamiltonian evolution. We have two options of transferring the polygon transformation to a Hamiltonian evolution: Either, we consider the transformation as a Hamiltonian and consider its usual Hamiltonian evolution, or, we consider each small iteration of the transformation as a new Hamiltonian and analyze how the eigenvalues and 
eigenstates change as in \cite{Unanayan2007}. Just like in \cite{Unanayan2007} we may assume that the the eigenstates are being kept fixed, while the the eigenvalues change. We can ``see'' what effect the geometric transformation has on the quantum system by looking at its Hamiltonian. In particular, we should be able to choose a Hamiltonian evolution for a desired eigenstate guided by some geometric transformation.

The results of both \cite{Unanayan2007} and \cite{VartziotisWipper2010} can be generalized to circulant Hermitian polygons in general. It therefore seems like we should be able to study the dynamics of the eigenstates for circular quantum systems.

\section{Polygon transformations based on similar triangles}
\label{sec:transformation}

\begin{figure}[htb]
  \centering
  \includegraphics[width=.9\linewidth]{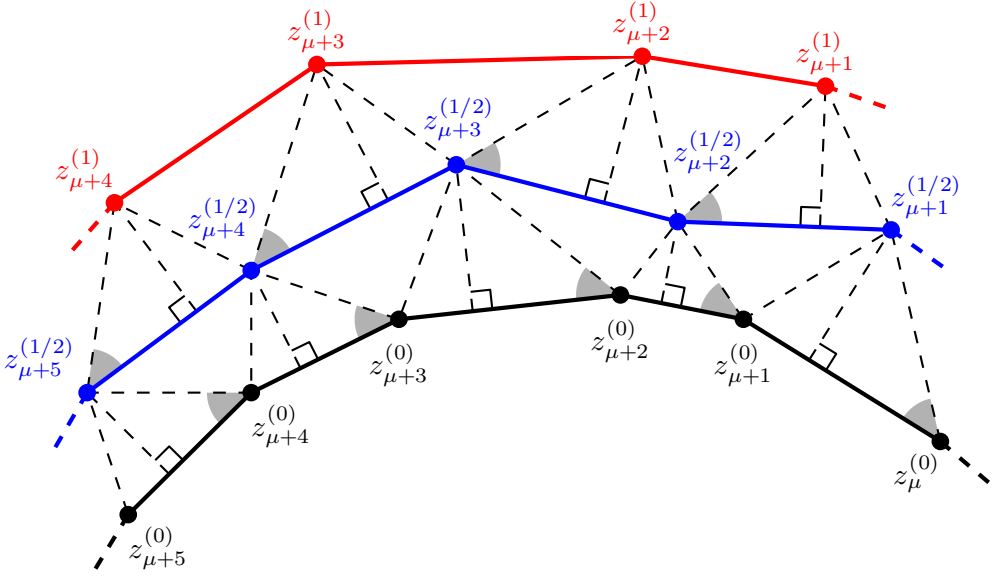}
  \caption{Initial polygon $z^{(0)}$ (black), $G_{-}$ transformed polygon
    $z^{(1/2)}=M_{-}z^{(0)}$ (blue), and $G_{+}$ transformed polygon
    $z^{(1)}=M_{+}z^{(1/2)}=Mz^{(0)}$ (red).}
  \label{fig:construction}
\end{figure}

Let $z^{(0)}=(z_0^{(0)},\dots, z_{n-1}^{(0)})^{\operatorname{t}}\in
\mathbb{C}^n$ denote an arbitrary polygon in the complex plane with $n\geq 3$
vertices $z_\mu^{(0)}$ using zero-based indices $\mu \in\{0,\dots,n-1\}$ and
sides $z_{\mu}^{(0)}z_{(\mu+1)\operatorname{mod} n}^{(0)}$ oriented according
to the order of vertices given by the vector $z^{(0)}$. A polygon transformation $G$ and its matrix
representation $M$ was defined and analyzed in \cite{VartziotisWipper2010} by constructing equally oriented similar triangles on
each side and taking the apices of these triangles which leads to a new
polygon with $n$ vertices.  In
Figure~\ref{fig:construction} it directly maps the polygon $z^{(0)}$ marked
black to the polygon $z^{(1)}$ marked red.

\begin{definition}
  \label{def:transformation}
  For $\theta \in (0,\pi/2)$, $\lambda\in (0,1)$,  and
  \[
    w:=\lambda + \mathrm{i} (1-\lambda)\tan\theta    
  \]
  let $G$ denote the polygon transformation $z^{(1)}=Mz^{(0)}$ defined by the
  matrix
  \begin{equation}
    \label{eq:transformationmatrix}
    (M)_{\mu,\nu}:=
    \begin{cases}
      |1-w|^2+|w|^2 & \text{if } \mu=\nu \\
      w(1-\overline{w}) & \text{if } \mu=(\nu+1)\operatorname{mod} n \\
      \overline{w}(1-w) & \text{if } \nu=(\mu+1)\operatorname{mod} n \\
      0 & \text{otherwise}
    \end{cases}\,,
  \end{equation}
  where $\mu,\nu\in\{0,\dots,n-1\}$. 
\end{definition}

\begin{figure}[htb]
  \hspace{2em}\includegraphics[width=.81\linewidth,clip=true]{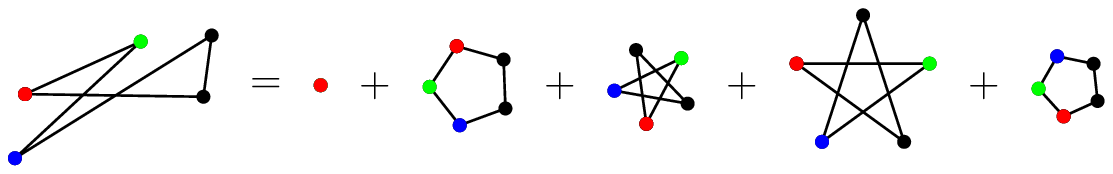}
  \medskip

  \hspace{2em}\includegraphics[width=.86\linewidth,clip=true]{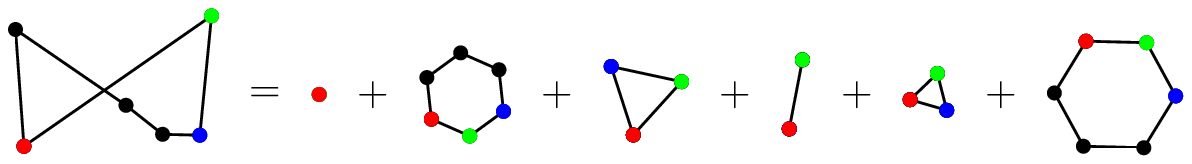}
  \caption{Decomposition of a random 5-gon (upper) and 6-gon (lower) into its eigenpolygons.}
  \label{fig:decomposition}
\end{figure}

It turns out that $M$ is a circulant Hermitian matrix. Furthermore, each row
and column of the matrices $M$ sum up to one, which
implies that it preserves the centroid,
\[
\frac{1}{n}\sum_{\mu=0}^{n-1}z_{\mu}^{(0)} =
\frac{1}{n}\sum_{\mu=0}^{n-1}z_{\mu}^{(1)}.
\]
It was shown in \cite{VartziotisWipper2010} that the eigenvalues of the transformation matrix $M$ for $n$-gons
  are positive 
  \begin{equation*}
    \label{eq:eigenvalues}
    \eta_k:=\big|1-\overline{w}+r^k\overline{w}\big|^2=|1-w|^2+|w|^2+2 \operatorname{Re}
    \big(r^k\overline{w}(1-w)\big) \,,
  \end{equation*}
  with $r:=\exp(2\pi\mathrm{i} /n)$ and $k\in\{0,\dots,n-1\}$. Figure~\ref{fig:decomposition} shows the
decomposition of random $n$-gons into eigenpolygons in the case of
$n\in\{5,6\}$. Here, the first three vertices have been colored red, green,
and blue respectively in order to denote the orientation. In particular, the left-most summand $n$ times the
centroid of the random $n$-gone, and the second summand is a counterclockwise oriented regular
$n$-gon.

\section{Quantum well in a circle}\label{sec:circle}

The solutions for one quantum well are of the form
\[
 \psi(x) =
 \begin{cases}
  A_1 \exp(\kappa x) & x \le -\frac{L}{2}\\
  A_{2}^+ \exp(-i{k} x) + A_{2}^- \exp(i{k} x)&  -\frac{L}{2} < x < \frac{L}{2}\\
  A_{3} \exp(-\kappa x)&  x \ge \frac{L}{2},
 \end{cases}
\]
where $\kappa$ and ${k}$ are chosen so that they satisfy the continuity conditions at $\pm\frac{L}{2}$ as well as the Schrödinger equation. If we have a quantum well in a circle of length $l = n \cdot a$, the solutions are of the form
\[
 \psi(x) =
 \begin{cases}
  \psi_1(x) = A_{1}^+ \exp(-\kappa x) + A_{1}^- \exp(\kappa (x - l))&   \frac{L}{2} \le x \le l - \frac{L}{2},\\
  \psi_2(x) = A_{2}^+ \exp(-i{k} x) + A_{2}^- \exp(i{k} x)&  -\frac{L}{2} < x < \frac{L}{2}\\
 \end{cases}
\]
where $x \in \R/\sim$ under the identification $x \sim x+l$ for $\kappa,{k}\in \R$, because the solutions are wave-like inside the quantum well and quickly decrease away from the quantum well.

Since the solutions inside the quantum well are either of the form cosine or sine
\[
 \psi(x) =
 \begin{cases}
  \psi_1(x) = A_{1}^+ \exp(-\kappa x) + A_{1}^- \exp(\kappa (x - l))&   \frac{L}{2} \le x \le l - \frac{L}{2},\\
  \psi_2(x) = A_{2} \cos({k} x) + A_{2}' \sin({k} x)&  -\frac{L}{2} < x < \frac{L}{2}\\
 \end{cases}
\]

We are looking for solutions $\psi$ to the Schrödinger equation
\begin{equation}\label{eq:schroedinger}
 \left\{T + V(x)\right\}\psi(x) = W \psi(x),
\end{equation}
where $T \equiv -\frac{\hbar^2}{2m}\frac{\partial^2}{\partial x^2}$ is the operator for the kinetic energy, 
\[
 V(x) = \begin{cases}
         0 & \frac{L}{2} \le x \le l - \frac{L}{2},\\
         -V_0 & -\frac{L}{2} < x < \frac{L}{2}
        \end{cases}
\]
is the operator for the potential energy, $-V_0 < W < 0$ is the potential energy of the electron represented by the wave function $\psi$, $\hbar$ is Planck's constant and $m$ is electron mass. Since
\[
-\frac{\hbar^2}{2m}\frac{\partial^2}{\partial x^2} \left(A_{1}^+ \exp(-\kappa x) + A_{1}^- \exp(\kappa (x - l))\right) = -\frac{\hbar^2 \kappa^2}{2m} \left(A_1^+ \exp(-\kappa x)  +A_{1}^- \exp(\kappa (x - l))\right), 
\]
and 
\[
-\frac{\hbar^2}{2m}\frac{\partial^2}{\partial x^2} \left(   A_{2}^+ \exp(-i{k} x) + A_{2}^- \exp(i{k} x) \right)= \frac{\hbar^2{k}^2}{2m} \left(   A_{2}^+ \exp(-i{k} x) + A_{2}^- \exp(i{k} x) \right)
\]
it follows that for a solution $\psi$ we have the eigenvalues
\begin{align*}
 -\frac{\hbar^2 \kappa^2}{2m}  &= W \quad \text{for } \frac{L}{2} \le x \le l - \frac{L}{2} \text{ and}\\
 \frac{\hbar^2{k}^2}{2m} - V_0 &= W \quad \text{for } -\frac{L}{2} < x < \frac{L}{2}
\end{align*}
and therefore
\begin{equation}\label{eq:klrel2}
\begin{split}
 {k} & = \sqrt{\frac{2m (W+V_0)}{\hbar^2}} \quad \text{for } -\frac{L}{2} < x < \frac{L}{2} \text{ and}\\
 \kappa & = \sqrt{-\frac{2m W}{\hbar^2}} \quad \text{for } \frac{L}{2} \le x \le l - \frac{L}{2} 
\end{split}
\end{equation}
This yields the equation
\[
 \kappa^2 + {k}^2 = \frac{2mV_0}{\hbar^2} \equiv C_0.
\]

We need the solutions to satisfy the continuity conditions
\begin{align*}
 \psi_1(l-\frac{L}{2}) &=  \psi_{2}(-\frac{L}{2}),   &\psi'_1(l-\frac{L}{2}) &=  \psi'_{2}(-\frac{L}{2}),\\
 \psi_1(\frac{L}{2}) &=  \psi_{2}(\frac{L}{2}),  &\psi'_1(\frac{L}{2}) &=  \psi'_{2}(\frac{L}{2}).
\end{align*}
We get the system of equations
\begin{equation*}
\begin{split} 
   A_1^+  \exp(- \kappa (l-\frac{L}{2})) + A_1^-  \exp(-\kappa \frac{L}{2})& = A_{2}^+ \exp(i{k} \frac{L}{2}) + A_{2}^- \exp(- i{k} \frac{L}{2})\\
   -\kappa A_1^+ \exp(- \kappa (l-\frac{L}{2})) +\kappa A_1^-  \exp( -\kappa \frac{L}{2})& = -i{k} A_{2}^+ \exp(i{k} \frac{L}{2}) +i{k} A_{2}^- \exp(- i{k} \frac{L}{2})\\
   A_1^+ \exp(-\kappa \frac{L}{2}) + A_1^-  \exp( \kappa (\frac{L}{2}-l))& = A_{2}^+ \exp(-i{k}\frac{L}{2}) + A_{2}^- \exp(i{k} \frac{L}{2})\\
   -\kappa A_1^+ \exp(-\kappa \frac{L}{2}) + \kappa A_1^-  \exp( \kappa (\frac{L}{2}-l)) & = -i{k} A_{2}^+ \exp(-i{k} \frac{L}{2}) +i{k} A_{2}^- \exp(i{k} \frac{L}{2}).
\end{split} 
\end{equation*}
Furthermore the expressions on the right side are really linear combinations of regular sine and cosine functions. We therefore get the following system of equations
\begin{equation*}
\begin{split} 
   A_1^+  \exp(- \kappa (l-\frac{L}{2})) + A_1^-  \exp(-\kappa \frac{L}{2})& = A_{2} \cos(-{k} \frac{L}{2}) + A_{2}' \sin(-{k} \frac{L}{2})\\
   -\kappa A_1^+ \exp(- \kappa (l-\frac{L}{2})) +\kappa A_1^-  \exp( -\kappa \frac{L}{2})& = -{k} A_{2} \sin(-{k} \frac{L}{2}) +{k} A_{2}' \cos(- {k} \frac{L}{2})\\
   A_1^+ \exp(-\kappa \frac{L}{2}) + A_1^-  \exp( \kappa (\frac{L}{2}-l))& = A_{2} \cos({k}\frac{L}{2}) + A_{2}' \sin({k} \frac{L}{2})\\
   -\kappa A_1^+ \exp(-\kappa \frac{L}{2}) + \kappa A_1^-  \exp( \kappa (\frac{L}{2}-l)) & = -{k} A_{2} \sin({k} \frac{L}{2}) +{k} A_{2}' \cos({k} \frac{L}{2}).
\end{split} 
\end{equation*}

In order to determine which $\kappa$ and ${k}$ satisfy the continuity condition, we need to find the values for which the coefficient matrix of the above system vanishes, i.e.
\[
\left|
\begin{matrix}
 \exp(- \kappa (l-\frac{L}{2})) & \exp(-\kappa \frac{L}{2}) & \cos(-{k} \frac{L}{2}) & \sin(-{k} \frac{L}{2})\\
 -\kappa\exp(- \kappa (l-\frac{L}{2})) & \kappa \exp( -\kappa \frac{L}{2}) & -{k}\sin(-{k} \frac{L}{2}) & {k} \cos(- {k} \frac{L}{2})\\
    \exp(-\kappa \frac{L}{2}) & \exp( \kappa (\frac{L}{2}-l))& \cos({k}\frac{L}{2}) & \sin({k} \frac{L}{2})\\
   -\kappa \exp(-\kappa \frac{L}{2}) & \kappa \exp( \kappa (\frac{L}{2}-l)) & -{k}\sin({k} \frac{L}{2}) &{k} \cos({k} \frac{L}{2})
\end{matrix} 
\right|= 0.
\]
Then we have
\begin{align*}
& \exp(−\kappa L−2 \kappa l) ((4 \kappa {k} \exp(\kappa L+\kappa l)+2 \kappa {k} \exp(2 \kappa L)+2 \kappa {k} \exp(2 \kappa l)) \sin(({k} L)/2)^2\\
& {}+((2 \kappa^2−2 {k}^2) \exp(2 \kappa L)+(2 {k}^2−2 \kappa^2) \exp(2 \kappa l)) \cos(({k} L)/2) \sin(({k} L)/2)\\
& {}+(4 \kappa {k} \exp(\kappa L+\kappa l)−2 \kappa {k} \exp(2 \kappa L)−2 \kappa {k} \exp(2 \kappa l)) \cos(({k} L)/2)^2)= 0
\end{align*}
A computation of $\kappa$ (and ${k}$) allows us to write down a continuously differentiable symmetric solution
\[
 \psi(x) =
 \begin{cases}
 \displaystyle
  \begin{split}\psi_1(x) = &A\frac{\cos({k} L/2)}{\exp(\kappa(L-l)+1)}\\
  &\cdot \left(\exp(-\kappa (x-L/2)) + \exp(\kappa (x - l+L/2))\right)\end{split} &   \frac{L}{2} \le x \le l - \frac{L}{2},\\
  \psi_2(x) = A\cos({k} x) &  -\frac{L}{2} < x < \frac{L}{2}\\
 \end{cases}
\]

\section{Constants in quantum mechanics}

In order to actually evaluate the above equations, various constants have to be set. Primarily 
\[
 C_0  \equiv \frac{2mV_0}{\hbar^2}.
\]

With the electron mass being $m=9.109 \cdot 10^{-31} kg$ and $\hbar = 1.055 \cdot 10^{-34} J \cdot s$ we get 

\[
 C_0  \equiv \frac{2 \cdot  9.109 \cdot 10^{-31} kg \cdot V_0}{(1.055 \cdot 10^{-34} J \cdot s)^2} \equiv 1.6368 \cdot 10^{38} \cdot V_0 \cdot \frac{s^2}{{kg}^2 \cdot m^4}.
\]

Changing from J to eV as the unit of energy, 
\[
 \frac{s^2}{kg \cdot m^2}  \equiv 1,602 \cdot 10^{-19} \cdot \frac{1}{eV}.
\]
we get

\[
 C_0  \equiv 2,6221 \cdot 10^{19} \cdot V_0 \frac{1}{eV \dot m^2}.
\]

Changing units to meV and nm we arrive at

\[
 C_0  \equiv 2,6221 \cdot 10^{-2} \cdot V_0 \frac{1}{meV \cdot {nm}^2}.
\]

\section{A circular quantum system}

We consider a quantum system given by $n$ quantum wells, which are of the same size and equidistantly distributed along a circle. The unperturbed problem follows the time-independent Schrödinger equation (as in Equation \eqref{eq:schroedinger})
\[
H\psi_\nu = W_\nu \psi_\nu
\]
where $H = T + V_\nu$ is the Hamilton operator, $T= \frac{d^2}{dx^2}$ and $V_\nu$ the potential function of the $\nu$-th quantum well.

For each individual quantum well we consider the symmetric solution $\psi_\nu$ with the lowest energy, and we choose a wave function $\Psi$ as a superposition of the solutions to the Schrödinger equation for the individual quantum wells
\begin{equation*}
\Psi = \sum\limits_\nu a_\nu \psi_\nu.
\end{equation*}

The Schrödinger equation for the quantum system with Energy $E$ therefore assumes the form
\begin{equation*}
H\Psi = E \Psi, \quad \text{where } H = T + \sum\limits_\nu V_\nu.
\end{equation*}
We apply perturbation theory to
\begin{equation*}
H \sum\limits_\nu a_\nu \psi_\nu = E  \sum_\nu a_\nu \psi_\nu,
\end{equation*}
multiply a basis function $\psi_\mu$ to the equation from the left and integrate over the circle
\begin{equation*}
\int\psi_\mu^* H \sum\limits_\nu a_\nu \psi_\nu = E \int \psi_\mu^* \sum\limits_\nu a_\nu \psi_\nu
\end{equation*}
and compute
\begin{equation*}
\sum\limits_\nu a_\nu \int\psi_\mu^* H \psi_\nu = E \sum\limits_\nu a_\nu  \int \psi_\mu^* \psi_\nu.
\end{equation*}
Notice that the wave functions $\psi_\mu$ are not orthogonal to each other. Therefore we define $n\times n$ matrices $H$ and $S$ with complex coefficients 
\begin{equation*}
H_{\mu\nu} = \int\psi_\mu^* H \psi_\nu  \quad \text{and} \quad S_{\mu\nu} = \int\psi_\mu^* \psi_\nu.
\end{equation*}
We therefore have
\begin{equation*}
\sum_\nu H_{\mu\nu} a_\nu  = E \sum\limits_\nu S_{\mu\nu} a_\nu.
\end{equation*}
We can rewrite this equation as a generalized eigenvalue problem
\begin{equation}\label{eq:generalized_eigenvalue_problem}
H a = E S a
\end{equation}
Clearly, $H$ and $S$ are Hermitian matrices. Its solution $a = (a_\nu)_{1,\ldots,n}$ and $E \in \R$ provides a solution $\Psi = \sum_\nu a_\nu \psi_\nu$ to the Schrödinger equation with Energy $E$ associated with the quantum system.

\section{Equal well depths}

The function $M$ in Definition \ref{def:transformation} is circulant. In order to relate $M$ to the matrix $H$ associated to the quantum system, all quantum wells must therefore have the same depth. Furthermore, the off-diagonal entries are not real numbers, therefore we need to choose our eigenfunctions $\psi_\nu$ such that the off-diagonal entries are not real. Lastly, $M$ is tridiagonal, therefore we assume that only adjacent quantum wells are coupled.

In order to compute the generalized eigenvalues, it helps to find the eigenvalues of $S$ first. If all potentials $V_\nu$ are equal, then $S$ is a circulant matrix
\[
 \begin{bmatrix}
1&s_2&\cdots &s_{n-1} & s_{n}\\
s_n & 1& \cdots& s_{n-2} & s_{n-1}\\
\vdots & \vdots &\ddots &\vdots & \vdots \\
s_3& s_4 & \cdots & 1 & s_2\\
s_2& s_3 & \cdots & s_n & 1
\end{bmatrix}
\]
with the additional property that $s_{n-\nu} = s_{2 + \nu}$. The normalized eigenvectors of a circulant matrix are given by
\[
 v_j = \frac{1}{\sqrt{n}} (1, \omega_j, \omega_j^2, \ldots , \omega_j^{n-1})^T \quad \text{for } j=0,\ldots,n-1,
\]
where $\omega_j = \exp(\frac{2ij}{n})$ are the $n$-th roots of unity and $i$ is the imaginary unit. The vectors $v_j$ and $\overline v_j$ are both eigenvectors with the same eigenvalues. The corresponding eigenvalues are then given by
\[
 \lambda_j = 1 + s_{2}\omega_j + s_{3}\omega_j^2 + \ldots + s_{n}\omega_j^{n-1}.
\]
The eigenvalues are real, if the matrix is symmetric.
The diagonal eigenvalue matrix is therefore
\[
 \Lambda_S = \text{diag}(\lambda_1,\ldots,\lambda_n),
\]
and the unitary eigenvector matrix is
\[
 \Phi_S = (v_0,\ldots,v_{n-1}).
\]
We have
\[
 S\Phi_S = \Phi_S\Lambda_S, \quad \text{or } \Phi_S^{-1}S\Phi_S = \Lambda_S.
\]
With
\[
 \Phi_S' = \Phi_S \Lambda_S^{-1/2}
\]
we get
\[
 (\Phi_S')^* S \Phi_S' = I.
\]
($\Phi_S'$ is not unitary.)
The matrix $H' = (\Phi'_S)^*H \Phi'_S$ is diagonal, because $\Phi_H = \Phi_S$  due to both matrices being circulant and therefore \[H' = \Lambda_S^{-1/2} \Phi_S^* H \Phi_S \Lambda_S^{1/2} = \Lambda_S^{-1/2} \Lambda_H \Lambda_S^{1/2} = \Lambda_H \Lambda_S^{-1}.\] Therefore $\Phi_{H'} = I$ and $\Lambda_{H'} = \Lambda_H \Lambda_S^{-1}$. 
Then we get for $\Phi := \Phi_S \Lambda_S^{-1/2}$ (not unitary)
\begin{align*}
 \Phi^* H \Phi & = (\Phi_S \Lambda_S^{-1/2})^* H \Phi_S \Lambda_S^{-1/2} = \Phi_{H'}^* H' \Phi_{H'}  = \Lambda_{H'}\\
 \Phi^* S \Phi & = I.
\end{align*}
Right multiplying the second equation by $\Lambda := \Lambda_{H'}$ and left multiplying both by $(\Phi^*)^{-1}$ yields
\[
 H\Phi = S\Phi \Lambda.
\]
Therefore, $\Lambda$ and $\Phi$ are the eigenvalue and eigenvector matrices of the generalized eigenvalue problem. The columns of $\Phi$ and the entries of $\Lambda$ correspond to the eigenvectors $a$ and $E$ in Equation \eqref{eq:generalized_eigenvalue_problem}.

\section{Negative vs. positive eigenvalues}

In the current setup for the quantum system, the energies turn out to be negative, while the eigenvalues of $M$ in Definition \ref{def:transformation} are positive. While the energies can be changed by parameterizing the quantum wells differently, it is necessary for $M$ to have positive eigenvalues to study the dynamical behavior.

We will again compute the Eigenvalue functions just like in Section \ref{sec:circle}. This time, we add a translating constant $V'\ge V_0$ to the function. Consider a finite quantum well given as
\[
 V(x) = \begin{cases}
         V' & |x| > \frac{L}{2}\\
         V' -V_0& |x| \le \frac{L}{2}.
        \end{cases}
\]
The computation goes through just like before and we get a continuously differentiable symmetric solution
\[
 \psi(x) =
 \begin{cases}
 \displaystyle
  \psi_1(x) = \begin{split}&A\frac{\cos({k} L/2)}{\exp(\kappa(L-l)+1)}\\ &\cdot\left(\exp(-\kappa (x-L/2)) + \exp(\kappa (x - l+L/2))\right)\end{split} &   \frac{L}{2} \le x \le l - \frac{L}{2},\\
  \psi_2(x) = A\cos({k} x) &  -\frac{L}{2} < x < \frac{L}{2}\\
 \end{cases}
\]
with
\begin{equation}
\begin{split}
 {k} & = \sqrt{\frac{2m (W+V_0-V')}{\hbar^2}} \quad \text{for } -\frac{L}{2} < x < \frac{L}{2} \text{ and}\\
 \kappa & = \sqrt{-\frac{2m (W-V')}{\hbar^2}} \quad \text{for } \frac{L}{2} \le x \le l - \frac{L}{2}.
\end{split}
\end{equation}

\section{Relation between $M$ and $H$}

The matrix $M$ from Definition \ref{def:transformation}
are triagonal matrices $M$ with entry $w$ along the diagonal and $w(1-\bar w)$ and $\bar w(1-w)$ along the sub- and super-diagonal, respectively, where $w := \lambda + i(1-\lambda)\tan \theta$. The eigenvalues are always positive, so are the entries in the diagonal. We can shift the potential function so that $H$ has only positive eigenvalues as follows.

We must have $W_1 := |1-w|^2 + |w|^2 = H_{\nu\nu} = \int \psi_\nu^* H \psi_\nu$. We compute
\begin{align*}
|1-w|^2 + |w|^2 & = (1-\lambda)^2 + (1-\lambda)^2 \tan^2\theta + \lambda^2 + (1-\lambda)^2 \tan^2\theta\\
& =  (1-\lambda)^2 + 2 (1-\lambda)^2 \tan^2\theta  + \lambda^2.
\end{align*}
If $\int \psi_1^* \psi_1 = 1$ then we can shift the potential function by $T := W_1 - H_{11}$. Then (calling the new matrix again $H$) we have $H_{11} = W_1$ by construction.
Furthermore,
\begin{align*}
W_2 := w(1-\bar w) = \lambda + i(1-\lambda)\tan \theta - \lambda^2 - (1-\lambda)^2 \tan^2\theta.
\end{align*}

Given $H_{\nu\nu}$ we can therefore fix $\lambda$ (e.g. $\lambda = 0.5$) and compute
\[
\theta = \arctan\sqrt{\frac{H_{\nu\nu} - (1-\lambda)^2 - \lambda^2}{2(1-\lambda)^2}}.
\]
This gives us $w(1-\bar w)$. The basis vectors $\psi_1$ and $\psi_2$ given by the eigenfunctions of a single quantum well ($1$ and $2$ respectively) can be changed into $\psi_1' = \alpha \psi_1 + i \beta \psi_2$ as well as $\psi_2' = \alpha \psi_2 - i \beta \psi_1$ for $\alpha, \beta \in \R$. We have
\begin{align*}
H'_{12} =  \int {\psi'_1}^* H \psi'_2 & =  \alpha^2 H_{12} - \beta^2 H_{21} - \alpha\beta i (H_{11} + H_{22}), \text{ and}\\
H'_{21} =  \int {\psi'_2}^* H \psi'_1 & =  \alpha^2 H_{21} - \beta^2 H_{12} + \alpha\beta i (H_{11} + H_{22}).
\end{align*}
Since $H_{12} = \overline H_{21}$ we get
\[
 H'_{12} = \overline H'_{21}.
\]
Therefore $H'$ is Hermitian.

Since $H_{12} = H_{21}$ and $H_{11} = H_{22}$ are real numbers we can compute $\alpha$ and $\beta$ such that
$W_2 = H'_{21}$ via \[\alpha^2 - \beta^2 = \frac{\real(W_2)}{2H_{12}} \quad \text{and} \quad \alpha\beta = \frac{\im(W_2)}{2H_{11}}.\]
This gives
\[
 \alpha^4 - \frac{\real(W_2)}{2H_{12}}\alpha^2 - \left(\frac{\im(W_2)}{2H_{11}}\right)^2 = 0
\]
and therefore
\begin{equation}\label{eq:compute_alpha}
 \alpha = \sqrt{\frac{\real(W_2)}{4H_{12}} + \sqrt{\left(\frac{\real(W_2)}{4H_{12}}\right)^2 + \left(\frac{\im(W_2)}{2H_{11}}\right)^2}}.
\end{equation}
The other parameter $\beta$ can then be computed via
\begin{equation}\label{eq:compute_beta}
 \beta = \frac{\im(W_2)}{2H_{11}\alpha}
\end{equation}

Alternatively, let us fix parameters $\theta$ and $\lambda$ for the geometric transformation. Let us choose $\lambda = \frac{1}{2}$. Then there are $\lfloor n/2 \rfloor + 1$ dominant eigenvalues
  \begin{equation*}
    \eta_k:=\big|1-\overline{w}+r^k\overline{w}\big|^2=|1-w|^2+|w|^2+2 \operatorname{Re}
    \big(r^k\overline{w}(1-w)\big) \,,
  \end{equation*}
  with $k\in\{0,\dots,\lfloor n/2\rfloor -1\}$ in the intervals $\theta \in (\theta_{k-1},\theta_k)$ for $\theta_{-1} = 0$, $\theta_{\lfloor n/2 \rfloor} = \frac{\pi}{2}$ and
  \[
    \theta_k = \frac{\pi}{2n}(2k+1).
  \]
  In the case $n=6$ we have $\theta_0 = 0$, $\theta_1 = \frac{\pi}{10}$, $\theta_2 =  \frac{3\pi}{10}$, $\theta_3 =  \frac{\pi}{2}$. Then for $\theta = \frac{\pi}{5}, \frac{2\pi}{5}$ the dominant eigenvalues are $\eta_1, \eta_2$, respectively.
  
As before, we may assume $W_1 = H_{11}$ after shifting the potential function appropriately. We can compute $\alpha, \beta \in \R$ satisfying $W_2 = H_{12}$ as in Equations \eqref{eq:compute_alpha} and \eqref{eq:compute_beta}.

In the case $n=6$, $\lambda = 0.5$ and $\theta = \frac{2\pi}{5}$ we have
\begin{align*}
H_{11} & =    5.2361 \\
H_{12} & = -2.1180 + 1.5388i\\
T & = 799.95\\
\alpha & =    1.6013\\
\beta & =   -0.57434.
\end{align*}

This shows that we can find a basis which gives geometric transformations corresponding to the triagonal matrices $M$ parametrized by $\theta$ and $\lambda$. However, this basis will not be normalized.

\section{Summary and Outlook}

We have interpreted the geometric transformation for polygons from \cite{VartziotisWipper2010} as a Hamiltonian for a system of quantum wells. We have argued that the Hamiltonian evolution corresponding to the iterated transformation will transform an arbitrary linear combination of eigenstates into a desired eigenstate. It will be interesting to see whether there is a parametrized family of Hermitian matrices for which the correspondence to quantum systems are more natural.


\bibliographystyle{elsarticle-num}
\bibliography{literature}







\end{document}